\documentclass[reqno]{amsart}
\title{Energy, Forces, Fields and the Lorentz Force Formula}
\author{Artice M. Davis\\
Professor Emeritus\\
San Jose State University}
\usepackage{graphicx}
\usepackage{endnotes}
\let\footnote=\endnote
\begin{document}
\maketitle
\begin{abstract}
We apply a simple decomposition to the energy of a moving particle. Based on this decomposition, we identify the potential and kinetic energies, then use them to give general definitions of momentum and the various kinds of forces exerted on the particle by fields, followed by the generalization of Newton's second law to accomodate these generally defined forces. We show that our generalization implies the Lorentz force law as well as Lagrange's equation, along with the usually accepted Lagrangian and the associated velocity dependent potential of a moving charged particle.
\end{abstract}
\subsection*{Introduction}The motivation for this paper is to present a rigorous derivation of the Lorentz force law in a nonrelativistic context. The provenance of the law is somewhat obscure. Lorentz's original paper,\footnote{H. A. Lorentz, \textit{La Th\'eorie El\'ectromagn\'etique de Maxwell et son Application aux Corps Mouvants} (Leiden, E. J. Brill, 1892).}, written in French and apparently not translated into English, assumed the ether as a medium and contained a number of unwarranted assumptions and vague definitions. Lorentz apparently then rejected his own derivation, choosing in his later monograph ``The Theory of Electrons''\footnote{H. A. Lorentz, \textit{The Theory of Electrons and its Applications to the Phenomena of Light and Radiant Heat}, 2\textsuperscript{nd} ed, (Teubner, Leipzig, 1916).} to simply say the law was ``...got by generalizing the results of electromagnetic experiments."\footnote{See the sentence immediately after equation (23), page 14 in reference [2]} He did not specify which experiments, but he clearly saw fit not to refer to his own earlier paper. Others have derived the formula by assuming a generalized Lagrangian for a moving charged particle,\footnote{See, for example, J. R. Taylor, \textit{Classical Mechanics}, (University Science Books, 2005), page 273; Goldstein, Poole, and Safco, \textit{Classical Mechanics}, (Addison-Wesley, San Francisco,2002), sec. 1.5, or M. G. Calkin, \textit{Lagrangian and Hamiltonian Mechanics},(World Scientific, Singapore, 1996), p. 46.} while still others have derived the Lagrangian presuming the Lorentz force formula.\footnote{O. D. Johns, \textit{Analytical Mechanics for Relativity and Quantum Mechanics}, (Oxford University Press, 2005), sec. 2.17. See also E. J. Konopinski, ``What the Electromagnetic Vector Potential Describes,'' Am. J. Phys., \textbf{46}, 499-502 (1978).} This type of analysis results in a generalized potential, namely \(\psi=\phi-\vec{A}\cdot\vec{v}\), for the moving particle, which we will derive. The quantity \(\vec{A}\) is the vector potential, which some authors (including Maxwell) referred to as the electromagnetic momentum; but others insist that the electromagnetic momentum is \(\epsilon_0\left[\vec{E}\times\vec{B}\right]\).\footnote{See D. Griffiths, ``Resource Letter EM-1: Electromagnetic Momentum,'' Am. J. Phys, \textbf{80}, 7-18, (2012) for a thorough discussion of this issue as well as for an extensive list of references.} We feel that much of the controversy in this matter, as well as in many others, is due to the lack of general definitions. We believe we have supplied such definitions in this work.\footnote{Our definitions do not generally resolve the aforementioned bone of contention, though it does imply that \(\vec{A}\) is the appropriate quantity insofar as the motion of a single charged particle is concerned.}

The work presented here assumes that the energy of a moving particle is a known quantity. The difficulty of defining energy has been discussed by others,\footnote{See, for example, Feynman, Leighton, and Sands, \textit{The Feynman Lectures on Physics}, (Addison-Wesley, Reading, 1964), v1, sec. 4.1. But see Falk, Hermann, and Schmid, ``Energy Forms or Energy Carriers?", Am. J. Phys., \textbf{51}, 1074-1076, (1983).} but we will deem it to be a primitive notion. Otherwise, we will adopt the operational point of view.\footnote{P. Bridgman, \textit{The Logic of Modern Physics}, (McMillan, New York, 1958), chapter 1.} This means that each fundamental quantity is defined by the description of a measuring instrument and a recipe for its use to measure the associated variable, while a derived quantity is defined by an equation expressing it in terms of previously defined fundamental quantities and/or previously defined derived quantities. We require, of course, that the latter never be self referential; that is, a flow diagram of all the definitions should contain no loops\textemdash it must be a tree structure.
\subsection*{Particles: Mass and Accelerational Force} We assume the usual definition of a particle, namely a vanishingly small region of space having certain properties through which it interacts with other similarly defined regions of space. We assume that such interactions are mediated by energy transfer through the empty spaces between them, these interactions depending upon the parameters mass and charge. Let us select two particles, remove them from any outside influence, place them at a given location in space with zero velocity, and measure their accelerations due to mutual influence.\footnote{Because their velocities are initially zero we are defining rest mass and ignoring relativistic effects. We will also assume ``near'' interactions so that retardation effects may be neglected} We will take it as a fact that in any such test the accelerations are always oppositely directed; furthermore, that the ratio of the magnitudes of the accelerations is always the same. Selecting one of these particles as a reference and denoting its acceleration by \(\vec{a}_0\) and that of the other by \(\vec{a}\), \footnote{We will assume throughout the following work that a vector can be thought of as a \(3\times 1\) column matrix. Hence, we can use matrix manipulations where needed.} we define the mass of the other particle by
\begin{equation}\label{defn of mass}
m=\frac{a_0}{a},
\end{equation}
where \(\mid\vec{a}\mid = a\) and \(\mid\vec{a}_0\mid=a_0\). If the two particles are identical then \(a_0=a\) so \(m=1\); hence, our reference particle has unit mass. Taking note of the opposing directions of the accelerations caused by the interaction, we can write
\begin{equation}\label{defn of accelerational force}
\vec{f}_a=-\vec{a}_0=m\vec{a},
\end{equation}
which we will define to be the accelerational force on the nonreference particle. This procedure for defining mass and force is due to Mach.\footnote{W. Mach, \textit{The Science of Mechanics}, (The Open Court Publishing Company, 1983) 4\textsuperscript{th} ed., chapter 2, section 5. There are additional consistency requirements which were pointed out by others after Mach advanced this definition. See L. Eisenbud, ``On the Classical Laws of Motion,'' Am. J. Phys., \textbf{26}, 144-159, (2005).} There exist forces, however, that are not associated with a moving object, for example the force of a spring on a weight it supports. We will offer more general definitions in a subsequent section of this paper.
\subsection*{Charge}Now let's select an arbitrary particle and test it against all the other particles in the universe, measuring the force between each pair. If no other particle repels it we will say it is uncharged. If, on the other hand, at least one other particle repels it, we will say our original particle is positively charged and assign it a charge of one unit thus making it our reference charge. Now we segregrate all particles in the universe into two classes: each of those in the first class repels our reference charge and is said to have a positive charge and each of those in the second class attracts the particle having our reference charge. Next, divide the second class into two subcategories: those which repel any other particle in the second class and those which attract each other particle in the second class. We will say that those in the repelling subcategory are negatively charged and those in the attracting subcategory uncharged. Thus, each and every particle in the universe is positively charged, negatively charged, or uncharged.\footnote{Note that the uncharged particles attract every other particle (weakly!) because of gravitational forces.} Finally, we invoke the Coulomb law to determine the magnitude of a given charge. Thus, we have defined charge, like mass, in an operational manner.

\subsection*{Energy Considerations}Consider a particle moving freely through space\textemdash that is, a particle with no mechanical constraints\textemdash and let \(U(\vec{r},\vec{v},t)\) be its energy.\footnote{We are assuming the particle has no energy of rotation (so no explicit dependence upon rotational angles or velocities).} Write
\begin{equation}\label{decomp1}
U(\vec{r},\vec{v},t)=U(\vec{r},0,t)+\left[U(\vec{r},\vec{v},t)-U(\vec{r},0,t)\right]=\phi(\vec{r},t)+T(\vec{r},\vec{v},t),
\end{equation}
where \(\phi\) and \(T\) have obvious definitions. The former is called the potential energy and the latter the kinetic energy. We define the generalized momentum by
\begin{equation}\label{momentum defn}
\vec{p}(\vec{r},\vec{v},t)=\nabla_{\vec{v}}T(\vec{r},\vec{v},t)=\nabla_{\vec{v}}U(\vec{r},\vec{v},t),
\end{equation}
where the subscript \(\vec{v}\) on the gradient operator refers to differentiation with respect to the components of velocity:
\begin{equation}\label{velocity gradient}
\nabla_{\vec{v}}=\hat{e}_i\frac{\partial}{\partial v_i}.
\end{equation}
Then we have
\begin{equation}\label{T integral}
T(\vec{r},\vec{v},t)=\int_0^{\vec{v}}\vec{p}(\vec{r},\vec{\alpha},t)\cdot d\vec{\alpha},
\end{equation}
where the integration is that of a line integral in velocity space with both position and time held fixed. Next, let us apply a similar decomposition to the generalized momentum, writing
\begin{equation}\label{decomp2}
\vec{p}(\vec{r},\vec{v},t)=\vec{p}(\vec{r},0,t)+\left[\vec{p}(\vec{r},\vec{v},t)-\vec{p}(\vec{r},0,t)\right]=\vec{A}(\vec{r},t)+\vec{Q}(\vec{r},\vec{v},t),
\end{equation}
where \(\vec{A}\) and \(\vec{Q}\) have obvious definitions. We will call \(\vec{A}\) the potential momentum due to the field and \(\vec{Q}\) the inertial momentum. Letting \(\vec{Q}=Q_j\hat{e}_j,\) define
\begin{equation}
\vec{m}_j(\vec{r},\vec{v},t)=\nabla_{\vec{v}}Q_j(\vec{r},\vec{v},t)=\hat{e}_i\frac{\partial Q_j}{\partial v_i}=m_{ij}\hat{e}_i.
\end{equation}
Then
\begin{equation}
Q_j(\vec{r},\vec{v},t)=\int_0^{\vec{v}}\vec{m}_j(\vec{r},\vec{\alpha},t)\cdot d\vec{\alpha}=\int_0^{\vec{v}}d\vec{\alpha}^T\vec{m}_j=\int_0^{\vec{v}}m_{ij}d\alpha_i.
\end{equation}
Finally, define the matrix
\begin{equation}
M=[m_{ij}]=\left[\frac{\partial Q_j}{\partial v_i}\right].
\end{equation}
We will call \(M=M(\vec{r},\vec{v},t)\) the generalized mass tensor. Using it, we have
\begin{equation}
\vec{Q}(\vec{r},\vec{v},t)=\int_0^{\vec{v}}d\vec{\alpha}^TM(\vec{r},\vec{\alpha},t),
\end{equation}
where \(d\vec{\alpha}^T\) is the transpose of the differential of the integration variable and \(d\vec{\alpha}^TM\) denotes the matrix product of the row matrix \(d\vec{\alpha}^T\) and the square matrix \(M\).\footnote{We are mixing matrix and vector notation here, but as all our vectors are column matrices this should not create confusion.}
\subsection*{The Classical Case}Classically, the mass tensor becomes
\begin{equation}
M=mI_3=m\begin{bmatrix}
1&0&0\\
0&1&0\\
0&0&1
\end{bmatrix},
\end{equation}
where \(m\) is a constant which we have already operationally defined.
Then the particle momentum is
\begin{equation}
\vec{Q}(\vec{r},\vec{v},t)=m\vec{v},
\end{equation}
the kinetic energy is
\begin{equation}
T(\vec{r},\vec{v},t)=\vec{A}(\vec{r},t)\cdot\vec{v}+\frac{1}{2}mv^2,
\end{equation}
and the total particle energy is
\begin{equation}\label{classical decomposition}
U(\vec{r},\vec{v},t)=\phi(\vec{r},t)+T(\vec{r},\vec{v},t)=\phi(\vec{r},t)+\vec{A}(\vec{r},t)\cdot\vec{v}+\frac{1}{2}mv^2.
\end{equation}
Following Maxwell, we will call the term \(\vec{A}\cdot\vec{v}\) the electrokinetic energy. In what follows, we will restrict ourselves to the classical case.
\subsection*{Generalized Forces}We will now define three generalized forces. The first will be ``positional force,'' the force on the particle caused by change of position. It will be defined by
\begin{equation}\label{positional force}
\vec{f}_P=-\nabla_{\vec{r}}\left[\phi(\vec{r},t)-\vec{A}(\vec{r},t)\cdot\vec{v}\right],
\end{equation}
where \(\nabla_{\vec{r}}=\hat{e}_i\partial/\partial x_i\). Here are the reasons for our choice of signs. If the field exerts force on the particle, the particle moves from a region of higher potential energy to a region of lower potential energy. On the other hand, a force exerted by the field on a particle tends to increase its kinetic energy\textemdash and \(\vec{A}\cdot\vec{v}\) is part of the kinetic energy.

The second generalized force we will define is the ``inertial force,'' given by the time rate of change of the generalized momentum:
\begin{equation}\label{inertial force}
\vec{f}_I(\vec{r},\vec{v},t)=\frac{d}{dt}\vec{p}(\vec{r},\vec{v},t)=\frac{d}{dt}\vec{A}(\vec{r},t)+\frac{d}{dt}Q(\vec{r},\vec{v},t)=\frac{d}{dt}\vec{A}(\vec{r},t)+m\vec{a}.
\end{equation}
We now generalize Newton's second law by postulating that
\begin{equation}\label{basic form of newton 2}
\vec{f}_P=\vec{f}_I.
\end{equation}
Supressing arguments for simplicity of notation and applying standard vector identities to equations \eqref{positional force}, \eqref{inertial force}, and \eqref{basic form of newton 2} , we obtain
\begin{equation}\label{second law}
-\nabla_{\vec{r}}\left[\phi-\vec{A}\cdot\vec{v}\right]=\nabla_{\vec{r}}\left[\vec{A}\cdot\vec{v}\right]-\vec{v}\times\left[\nabla_{\vec{r}}\times\vec{A}\right]+\partial_t\vec{A}+m\vec{a}.
\end{equation}
Our third and last generalized force is that part of the inertial force which we have already defined to be the '`accelerational force,'' \(\vec{f}_a=m\vec{a}\). Using it in equation \eqref{second law} gives
\begin{equation}\label{second second law}
\vec{f}_a=-\nabla_{\vec{r}}\phi-\partial_t\vec{A}+\vec{v}\times\left[\nabla_{\vec{r}}\times\vec{A}\right].
\end{equation}
If we define the `positional field'' by
\begin{equation}
\vec{E}=-\nabla_{\vec{r}}\phi-\partial_t\vec{A}
\end{equation}
and the ``motional field'' by
\begin{equation}
\vec{B}=\nabla_{\vec{r}}\times\vec{A},
\end{equation}
we can rewrite equation \eqref{second second law} in the extremely simple form
\begin{equation}\label{field force}
\vec{f}_a=\vec{E}+\vec{v}\times\vec{B}.
\end{equation}
We see at once that \(\vec{E}\) is the accelerational force on a particle at rest and \(\vec{v}\times\vec{B}\) the added accelerational force due to its motion. These interpretations clearly serve as operational definitions of these two fields.

\subsection*{The Lorentz Force Law}We now recognize that the energy of a particle might depend upon both mass and charge. At first suppressing the position, velocity, and time arguments and then reintroducing them, we write
\begin{equation}
U=U(m,q)=U(m,0)+\left[U(m,q)-U(m,0)\right]=V_m(\vec{r},\vec{v},t)+W_q(\vec{r},\vec{v},t),
\end{equation}
with obvious definitions of \(V_m\) and \(W_q\). Next, we perform, for each of \(V_m\) and \(W_q\) the decomposition in equation \eqref{classical decomposition}, using obvious notation and assuming that each component is normalized to unit mass or charge as appropriate:
\begin{equation}\label{V decomposition}
V_m(\vec{r},\vec{v},t)=m\phi_m(\vec{r},t)+\frac{1}{2}mv^2.
\end{equation}
and
\begin{equation}\label{W decomposition}
W_q(\vec{r},\vec{v},t)=q\phi_q(\vec{r},t)+q\vec{A}(\vec{r},t)\cdot\vec{v}.
\end{equation}
We have made two key assumptions here, namely
\begin{enumerate}
\item \(V_m\) has no component due to field momentum.
\item \(W_q\) is independent of particle mass.
\end{enumerate}
These assumptions can, of course, be removed at the expense of a more complex resulting theory. It is also convenient in working strictly with electrodynamics to assume that \(m\phi_m<<q\phi_q\approx q\phi\). These assumptions, taken together, permit us to remove all subscripts and rewrite the total particle energy in equation \eqref{classical decomposition} as
\begin{equation}\label{classical decomposition 2}
U(\vec{r},\vec{v},t)=q\phi(\vec{r},t)+q\vec{A}(\vec{r},t)\cdot\vec{v}+\frac{1}{2}mv^2
\end{equation}
and our field force equation in \eqref{field force} as
\begin{equation}\label{lorentz}
\vec{f}_a=q\vec{E}+q\vec{v}\times\vec{B},
\end{equation}
where all terms except \(\vec{f}_a\) are purely electrical in nature.
We now see that \(\vec{E}\) is clearly the electrical field intensity and \(\vec{B}\) the magnetic field as these quantities are normally defined in electromagnetic field theory, the former being the per-unit force on a stationary charged particle and the second term in equation \eqref{lorentz} being the incremental force added by the magnetic field.
\subsection*{Lagrange's Equation} Let's return to Newton's generalized second law, \(\vec{f}_p=\vec{f}_I\), and write it in terms of the basic definitions:
\begin{equation}\label{general newtons second law}
-\nabla_{\vec{r}}\left[q\phi-q\vec{A}\cdot\vec{v}\right]=\frac{d}{dt}\nabla_{\vec{v}}\left[q\vec{A}\cdot\vec{v}+\frac{1}{2}mv^2\right].
\end{equation}
Note that
\begin{equation}\label{defn of em T}
T=\frac{1}{2}mv^2+q\vec{A}\cdot\vec{v}.
\end{equation}
In component form \eqref{general newtons second law} becomes
\begin{equation}\label{lagrange components}
\frac{d}{dt}\frac{\partial T}{\partial v_i}+q\frac{\partial\phi}{\partial x_i}=\frac{\partial}{\partial x_i}\left[q\vec{A}\cdot\vec{v}\right].
\end{equation}
Next, we define
\begin{equation}\label{defn of L}
L=T-q\phi=\frac{1}{2}mv^2-q\left[\phi-\vec{A}\cdot\vec{v}\right]
\end{equation}
 and note that \(\phi\) is not a function of \(\vec{v}\) to obtain
\begin{equation}\label{lagranges equation}
\frac{d}{dt}\frac{\partial L}{\partial v_i}-\frac{\partial L}{\partial x_i}=0.
\end{equation}
We now see that
\begin{equation}\label{generalized potential}
\psi(\overline{r},\overline{v},t)=\phi(\overline{r},t)-\overline{A}(\overline{r},t)\cdot\vec{v}
\end{equation}
 has the character of a velocity dependent potential. We also see that \(L\) is the commonly accepted Lagrangian for a particle in the electromagnetic field, usually either assumed on an \textit{ad hoc} basis or derived from the Lorentz force law as an assuption.\footnote{See the paper by Konopinski in reference [5].} Equation \eqref{lagranges equation} can, of course, be immediately extended by expressing the cartesian position variables in terms of generalized coordinates. Standard procedures can then be applied to show that Lagrange's equation is invariant under this transformation. Thus, the theory just outlined produces both the Lorentz force equation and Lagrange's equation as results, rather than as assumptions.
 
 Finally, we note that the steps leading from the generalized Newton's second law expresssd by equation \eqref{basic form of newton 2} with its associated force definitions to Lagrange's equation in \eqref{lagranges equation} are all reversible; in other words the two equations are equivalent mathematical assertions. Hence, if one accepts the latter, one must accept the former. We feel, therefore, that the theory outlined in this paper offers a solid, theoretically sound approach to introducing both topics in introductory courses in fields and classical mechanics.
 \subsection*{Acknowledgements} I would like to express my appreciation to Vladimir Onoochin for a number of productive discussions about the topics treated her.
\theendnotes 
\end{document}